\documentstyle[preprint,aps,eqsecnum]{revtex}
\begin{document}

\draft
\title{Inflation at Low Scales: General Analysis and a Detailed Model}
\author{William H. Kinney\thanks{Electronic address: kinney@colorado.edu} and
K.T. Mahanthappa\thanks{Electronic address: ktm@verb.colorado.edu}}
\address{Dept of Physics, Campus Box 390}
\address{University of Colorado, Boulder CO 80309-0390}
\author{COLO-HEP-366, hep-ph/9512241 (Submitted to Physical Review D)}
\date{December 5, 1995}
\maketitle

\begin{abstract}
Models of inflationary cosmology based on spontaneous symmetry breaking
typically suffer from the shortcoming that the symmetry breaking scale is
driven to nearly the Planck scale by observational constraints.
In this paper we investigate inflationary potentials in a general context,
and show that this difficulty is characteristic only of potentials $V(\phi)$
dominated near their maxima by terms of order $\phi^2$. We find that
potentials dominated by terms of order $\phi^m$ with \hbox{$m > 2$} can
satisfy observational constraints at an arbitrary symmetry breaking scale. Of
particular interest, the spectral index of density fluctuations is shown to
depend only on the order of the lowest non-vanishing derivative of $V(\phi)$
near the maximum. This result is illustrated in the context of a specific
model, with a broken ${\rm SO(3)}$ symmetry, in which the potential is
generated by gauge boson loops.
\end{abstract}


\section {Introduction}
\label{secintro}
Inflation is an attractive model of the early universe because it naturally
explains the smoothness and flatness of the universe, and provides a
well-defined mechanism for the creation of primordial density fluctuations.
In a typical inflationary model, spontaneous symmetry breaking at some scale
$v$ creates a potential for a scalar field with nonzero vacuum energy density
$\Lambda^4$, resulting in an epoch in which the universe is dominated by
vacuum energy and undergoes a period of exponential expansion.
Quantum fluctuations in the scalar field create density fluctuations which
later collapse to form galaxies and clusters of
galaxies\cite{hawking82,starob82,guth82,bardeen83}. Although the inflationary
paradigm is widely accepted, most attempts to incorporate inflation into
specific models of particle physics suffer from two shortcomings:
(1) Parameters such as coupling constants must often be ``fine-tuned'' to
extremely small values in order to avoid massive overproduction of density
fluctuations. This can be seen in a general sense to arise from an
exponential dependence of the vacuum energy density $\Lambda^4$ on the
symmetry breaking scale $v$.
(2) Symmetry breaking scales significantly below the Planck scale, \hbox{$v <
m_{Pl} \simeq 10^{19} {\rm GeV}$} are inconsistent with observational
constraints. The latter difficulty is especially troubling, since physics at
the Planck scale is currently poorly understood, and there is no compelling
reason to expect that standard notions of spontaneous symmetry breaking are
valid at such high energies.

Section \ref{secslowrollreview} of this paper provides a brief review of
slow-roll and inflationary constraints. In section \ref{secgeneralpotentials}
we show, using a general class of scalar field potentials, that the problems
of fine-tuning and of scales being driven close to $m_{Pl}$ are in fact
characteristic only of a restricted class of potentials, those which are
dominated by terms of order $\phi^2$ in the scalar field driving inflation.
We obtain the result that for potentials dominated by terms of order $\phi^m$
with \hbox{$m > 2$}, inflation can take place consistent with cosmological
constraints at an arbitrary symmetry breaking scale. Of particular interest,
we show that for \hbox{$m > 2$}, the spectral index $n_s$ of the scalar
density fluctuations is {\it independent} of the specific form of the
potential, and is determined entirely by the order $m$ of the lowest
non-vanishing derivative at the maximum of the potential, with \hbox{$0.93 <
n_s < 0.97$} for all orders $m$. The results of this section have been
briefly reported earlier \cite{kinney952}. In section \ref{secso3model}, we
construct a specific model in which the inflationary potential is created by
radiative corrections from gauge bosons in a Lagrangian with an explicitly
broken ${\rm SO(3)}$ symmetry. This model is dominated by terms of order
$\phi^4$, and demonstrates the general result in a detailed context.

\section {Slow-Roll and Inflationary Constraints}
\label{secslowrollreview}

\subsection{Inflationary dynamics and constraints}
\label{secinflgeneral}
Inflationary cosmologies explain the observed flatness and homogeneity of the
universe by postulating the existence of an epoch during which the energy
density of the universe was dominated by vacuum energy, resulting in a period
of exponential increase in the scale factor of the
universe\cite{guth81,linde82,albrect82},
\begin{equation}
a(t) \propto e^{H t}.
\end{equation}
The {\it Hubble parameter} H is given by
\begin{equation}
H^2 \equiv \left({\dot a \over a}\right)^2 = {8 \pi \over 3 m_{Pl}^2}
\rho_{vac} \simeq const.,
\end{equation}
where \hbox{$m_{Pl} \simeq 10^{19} \mbox{GeV}$} is the Planck mass. Nonzero
vacuum energy is introduced into particle physics models by including a
scalar field $\phi$, the {\it inflaton}, with a potential
$V\left(\phi\right)$. During inflation, the inflaton is displaced from the
minimum of its potential, resulting in a nonzero vacuum energy, and evolves
to the minimum with equation of motion
\begin{equation}
\ddot\phi + 3 H \dot\phi + V'\left(\phi\right) = 0.
\end{equation}
So-called ``new'' inflationary models\cite{linde82,albrect82} consider
potentials which contain at least one region flat enough that the evolution
of the field is friction dominated, and the equation of motion can be taken
to be:
\begin{equation}
3 H \dot\phi + V'\left(\phi\right) = 0.\label{slowroll}
\end{equation}
This is known as the {\it slow-roll} approximation. This approximation can be
shown to be valid if the {\it slow-roll parameters} $\epsilon$ and
$\left|\eta\right|$\cite{copeland93} are both less than 1, where:
\begin{equation}
\epsilon\left(\phi\right) \equiv {m_{Pl}^2 \over 16 \pi}
\left({V'\left(\phi\right) \over V\left(\phi\right)}\right)^2,\end{equation}
and
\begin{equation}\eta\left(\phi\right) \equiv {m_{Pl}^2 \over 8 \pi}
\left[{V''\left(\phi\right) \over V\left(\phi\right)} - {1 \over 2}
\left({V'\left(\phi\right) \over V\left(\phi\right)}\right)^2\right].
\end{equation}
Consider a scalar field at an initial value $\phi$. The field evolves
according to equation (\ref{slowroll}) to the minimum of the potential, where
it oscillates and decays into other particles ({\it reheating}). Inflation
ends and reheating commences at a field value $\phi_f$ where the first order
parameter $\epsilon\left(\phi_f\right)$ is unity\cite{copeland93}:
\begin{equation}
\epsilon\left(\phi_f\right) \equiv {m_{Pl}^2 \over 16 \pi}
\left({V'\left(\phi_f\right) \over V\left(\phi_f\right)}\right)^2 = 1,
\end{equation}
where \hbox{$\epsilon\left(\phi\right) < 1$} during the inflationary period.
The number of e-folds of inflation which occur when the field evolves from
$\phi$ to $\phi_f$ is
\begin{equation}
N\left(\phi\right) = {8 \pi \over m_{Pl}^2}
\int_{\phi_f}^{\phi}{{V\left(\phi'\right) \over
V'\left(\phi'\right)}\,d\phi'}.\label{numefolds}
\end{equation}
Smoothness on scales comparable to the current horizon size requires \hbox{$N
\geq 60$}, which places an upper limit on the initial field value \hbox{$\phi
\leq \phi_{60}$}, where \hbox{$N\left(\phi_{60}\right) \equiv 60$}. Quantum
fluctuations in the inflaton field produce density fluctuations on scales of
current astrophysical interest when \hbox{$\phi \simeq \phi_{60}$}. The
scalar density fluctuation amplitude produced during inflation is given
by\cite{stewart93}:
\begin{eqnarray}
\delta =&& \left({2 \over \pi}\right)^{1/2}
{\left[V\left(\phi_{60}\right)\right]^{3/2} \over m_{Pl}^3
V'\left(\phi_{60}\right)} \left\lbrace1 - \epsilon\left(\phi_{60}\right) +
\left(2 - \ln2 - \gamma\right) \left[2 \epsilon\left(\phi_{60}\right) -
\eta\left(\phi_{60}\right)\right]\right\rbrace\cr
\simeq&& \left({2 \over \pi}\right)^{1/2}
{\left[V\left(\phi_{60}\right)\right]^{3/2} \over m_{Pl}^3
V'\left(\phi_{60}\right)},\label{deltainfl}
\end{eqnarray}
where \hbox{$\gamma \simeq 0.577$} is Euler's constant. The amplitude of
tensor, or gravitational wave, fluctuations is\cite{stewart93},
\begin{eqnarray}
a_T =&& {2 \over \sqrt{3 \pi}} {\left[V\left(\phi_{60}\right)\right]^{1/2}
\over m_{Pl}^2} \left[1 + \left(1 - \ln2 - \gamma\right)
\epsilon\left(\phi_{60}\right)\right]\cr
\simeq&& {2 \over \sqrt{3 \pi}} {\left[V\left(\phi_{60}\right)\right]^{1/2}
\over m_{Pl}^2}.\label{tensoramplitudedef}
\end{eqnarray}
In addition, it is possible to calculate the spectral index $n_s$ of the
scalar density fluctuations. The fluctuation power per logarithmic interval
$P\left(k\right)$ is defined in terms of the density fluctuation amplitude
$\delta_k$ on a scale $k$ as \hbox{$P\left(k\right) \equiv |\delta_k|^2$}.
The {\it spectral index} $n_s$ is defined by assuming a simple power-law
dependence of $P\left(k\right)$ on $k$, \hbox{$P\left(k\right) \propto
k^{n_s}$}.
The spectral index of density fluctuations, $n_s$, is given in terms of the
slow-roll parameters $\epsilon$ and $\eta$\cite{stewart93}:
\begin{equation}
n_s \simeq 1 - 4 \epsilon\left(\phi_{60}\right) + 2
\eta\left(\phi_{60}\right)\label{scalarindex}.
\end{equation}
For \hbox{$\epsilon,\,\left|\eta\right| \ll 1$} during inflation,
inflationary theories predict a nearly scale-invariant power spectrum,
\hbox{$n_s \simeq 1$}.

In order for an inflationary model to be viable, it must satisfy
observational constraints from the Cosmic Background Explorer (COBE)
Differential Microwave Radiometer (DMR) observation of the cosmic microwave
background (CMB) fluctuations\cite{smoot92,wright92}:
(1) The density fluctuation amplitude is observed to be \hbox{$\delta \simeq
10^{-5}$}. (2) The value of the spectral index derived from the first year of
COBE data is \hbox{$n_s = 1.1 \pm 0.5$}. (The COBE two year results are also
available\cite{bennett94}. However, different statistical methods used in
analyzing the data lead to different bounds on the spectral
index\cite{bennett95}. For the purposes of this paper, we will take
\hbox{$n_s \geq 0.6$}.)
For models in which the field can be taken to be initially at the maximum of
the potential, there is an additional consistency constraint. If the maximum
of the potential is an unstable equilibrium, a field sitting at the maximum
will be driven off by quantum fluctuations of amplitude $\phi_q$, where
$\phi_q$ on the scale of a horizon size is given by:
\begin{equation}
\phi_q = {H \over 2 \pi} = {1 \over 2 \pi} \sqrt{{8 \pi \over 3 m_{Pl}^2}
V\left(0\right)}.
\end{equation}
If \hbox{$N\left(\phi_q\right) < 60$}, the universe will not inflate
sufficiently, so we have the constraint for successful inflation that
\begin{equation}
\left(\phi_q / \phi_{60}\right) < 1.\label{quantumconstraint}
\end{equation}
In the case of models, such as natural inflation, in which the initial
conditions are randomly selected, this constraint takes a somewhat different
form.

\subsection{Natural inflation}
\label{secnaturalinflation}
``Natural'' inflationary theories\cite{freese90,adams93} use a pseudo
Nambu-Goldstone boson to drive the inflationary expansion. The basic scenario
consists of the following: A spontaneous symmetry breaking phase transition
occurs at a scale $v$, and temperature \hbox{$T \simeq v$}. For definiteness,
consider the simple case of symmetry breaking involving a single complex
scalar field $\phi$, with a potential
\begin{equation}
V\left(\phi\right) = \lambda \left[\phi^* \phi - v^2\right]^2,
\end{equation}
which is symmetric under a global ${\rm U(1)}$ transformation \hbox{$\phi
\rightarrow e^{i \alpha} \phi$}.
At the minimum of the potential $V\left(\phi\right)$, we can parameterize the
scalar field as \hbox{$\phi = \sigma \exp\left[{i \theta} / v\right]$}.
The radial field $\sigma$ has a mass \hbox{$M^2_\sigma \simeq \lambda v^2$}.
The field $\theta$ is a Nambu-Goldstone boson, and is massless at tree level.
If the ${\rm U(1)}$ symmetry of the potential $V\left(\theta\right)$ is
preserved by the rest of the Lagrangian, $\theta$ remains massless with loop
corrections. But if the ${\rm U(1)}$ is broken by other terms in the
Lagrangian, $\theta$ acquires a potential $V_1\left(\theta\right)$ from loop
corrections, and acquires a nonzero mass. Then $\theta$ is called a {\it
pseudo Nambu-Goldstone boson} (PNGB). (Models with explicitly broken global
${\rm U(1)}$ symmetries have been discussed in the
literature\cite{adams93,kinney95}. In this paper, we consider an inflationary
potential created by gauge boson loop effects in a Lagrangian with an
explicitly broken ${\rm SO(3)}$ symmetry.)
Assuming the mass of $\theta$ is much less than that of the radial mode,
\hbox{$M^2_\theta \ll M^2_\sigma$}, the field $\theta$ is effectively
massless near the original symmetry breaking scale \hbox{$T \simeq v$}. As
the temperature of the universe decreases, \hbox{$T \ll M_\sigma$},
excitations of the heavy $\sigma$ field are damped, so that we can take
\hbox{$\sigma = v = const.$} The only remaining degree of freedom is
$\theta$, and we can parameterize $\phi$ as:
\begin{equation}
\phi = v e^{i \theta / v}.\label{adiabatic}
\end{equation}
At temperatures \hbox{$T \gg M_\theta$}, the effective potential
$V_1\left(\theta\right)$ is negligible. When the universe cools to \hbox{$T
\simeq M_\theta$}, $V_1\left(\theta\right)$ becomes important\cite{freese90}.
The field $\theta$ rolls down the potential to its minimum, resulting in
inflationary expansion during the period in which the energy density of the
universe is dominated by vacuum energy. Natural inflationary models typically
assume a potential for the PNGB of the form
\begin{equation}
V_1\left(\theta\right) = \Lambda^4 \left[1 + \cos\left({\theta \over
v}\right)\right],\label{stdnat}
\end{equation}
where $v$ is the original symmetry breaking scale, and $\Lambda$ is an
independent energy scale characterizing the temperature at which the
potential $V_1\left(\theta\right)$ becomes significant.

Note that in natural inflation we {\it cannot} assume an initial field value
near the quantum fluctuation limit $\theta_q$, and the consistency condition
(\ref{quantumconstraint}) must be modified.
As the universe cools to \hbox{$T \simeq \Lambda$}, we expect the field
$\theta$ and its derivative $\dot\theta$ to take on different values in
different regions of the universe; we assume that the field is, to a good
approximation, uniform within any pre-inflation horizon volume. The universe
just prior to inflation therefore consists of a large number of causally
disconnected regions, each with independent initial conditions for $\theta$
and $\dot\theta$.
Each independent region will inflate a different amount, or perhaps not at
all, depending on the conditions within that region. In a successful model
for inflation, the {\it post}-inflation universe is strongly dominated by
regions in which \hbox{$N\left(\theta\right) \geq 60$}. It can be shown that
the initial value of $\dot\theta$ does not significantly affect the number of
e-folds of inflation\cite{knox93}, and hence we consider here only the upper
limit \hbox{$\theta \leq \theta_{60}$}. Consider a pre-inflation horizon
volume $V_0$ and initial field value $\theta$; during inflation, this region
expands to a volume \hbox{$V = V_0 \exp\left[3 N\left(\theta\right)\right]$}.
The fraction of the volume of the post-inflation universe for which
\hbox{$N\left(\theta\right) \geq 60$} is then \cite{adams93}
\begin{equation}
F\left(N \geq 60\right) = 1 -
\left(\int_{\theta_{60}}^{\theta_{min}}{\exp\left[3
N\left(\theta\right)\right]\,d\theta} \Bigg/
\int_{\theta_q}^{\theta_{min}}{\exp\left[3
N\left(\theta\right)\right]\,d\theta}\right),
\end{equation}
where $\theta_{min}$ is the minimum of the potential, and
\hbox{$N\left(\theta > \theta_f\right) \equiv 0$}. Here a cutoff
\hbox{$\theta_q \equiv H / 2 \pi$}, the magnitude of quantum fluctuations on
the scale of a horizon size, has been introduced as the lower limit for the
field value. A successful inflationary theory has the resulting
characteristic that
\begin{equation}
\Pi\left(\theta_{60}\right) \equiv
\int_{\theta_{60}}^{\theta_{min}}{\exp\left[3
N\left(\theta\right)\right]\,d\theta} \Bigg/
\int_{\theta_q}^{\theta_{min}}{\exp\left[3
N\left(\theta\right)\right]\,d\theta} \ll 1.\label{probparameter}
\end{equation}
This condition can be satisfied even for models in which $\theta_{60}$ is
constrained to be extremely small, as long as \hbox{$\theta_q \ll
\theta_{60}$}, and \hbox{$N\left(\theta_q\right) \gg  60$}. The integral
(\ref{probparameter}) is in most cases difficult to calculate. However, a
very rough limit on its magnitude can be taken to be:
\begin{equation}
\Pi < {\exp\left[3 N\left(\theta_{60}\right)\right] \over \exp\left[3
N\left(\theta_q\right)\right]}.
\end{equation}
In the cases considered in this paper, natural inflation models which satisfy
other observational constraints are also characterized by \hbox{$\Pi \ll 1$},
and this consistency condition does not provide a significant constraint.

\section{Inflationary Constraints for General Potentials}
\label{secgeneralpotentials}

Scalar field potentials arising from spontaneous symmetry breaking can in
general be characterized by the presence of a ``false'' vacuum, an unstable
or metastable equilibrium with nonzero vacuum energy density, and a physical
vacuum, for which the classical expectation value of the scalar field is
nonzero. At the physical vacuum, the potential has a stable minimum where the
vacuum energy density is defined to vanish. In this paper, we consider
potentials for ``new'' inflation, where the false vacuum is an {\it unstable}
equilibrium, and inflation takes place during a period of slow-roll. Take a
potential $V\left(\phi\right)$ described by a symmetry breaking scale $v$ and
a vacuum energy density $\Lambda^4$:
\begin{equation}
V\left(\phi\right) = \Lambda^4 f\left({\phi \over
v}\right).\label{dimensionlessf}
\end{equation}
We take the first derivative of the potential to be zero at the origin and
the minimum to be at \hbox{$\phi = \phi_{min} \sim v$}:
\begin{eqnarray}
&&f'\left(0\right) = f'\left(\phi_{min}\right) = 0,\cr
&&f\left(0\right) = 1,\ f\left(\phi_{min}\right) = 0.
\end{eqnarray}
The first order inflationary parameter $\epsilon\left(\phi\right)$ is given
by
\begin{eqnarray}
\epsilon\left(\phi\right) \equiv&& {m_{Pl}^2 \over 16 \pi}
\left({V'\left(\phi\right) \over V\left(\phi\right)}\right)^2\cr
=&& {1 \over 16 \pi} \left({m_{Pl} \over v}\right)^2 \left({f'\left(\phi /
v\right) \over f\left(\phi / v\right)}\right)^2.\label{genericepsilon}
\end{eqnarray}
An inflationary phase is characterized by \hbox{$\epsilon < 1$}: here
\hbox{$\epsilon\left(\phi = 0\right) = 0$} by construction. If
$\epsilon\left(\phi\right)$ is everywhere increasing on the range \hbox{$0
\leq \phi < \phi_{min}$}, there is a unique field value $\phi_f$ at which
inflation ends, where \hbox{$\epsilon\left(\phi_f\right) \equiv 1$} and
\hbox{$\epsilon\left(\phi < \phi_f\right) < 1$}.
We are particularly interested in cases where the symmetry breaking takes
place well below the Planck scale, \hbox{$v \ll m_{Pl}$}. Noting from
(\ref{genericepsilon}) that \hbox{$\epsilon \propto \left(m_{Pl} /
v\right)^2$}, the field value $\phi_f$ at which inflation ends is small for
\hbox{$v \ll m_{Pl}$}, and we need only consider the behavior of the
potential near the origin. We perform a Taylor expansion of
$V\left(\phi\right)$ about the origin:
\begin{equation}
V\left(\phi\right) = V\left(0\right) + {1 \over m!} {d^m V \over d \phi^m}
\biggr|_{\phi =  0} \phi^m + \cdots,\label{generalTaylorexpansion}
\end{equation}
where \hbox{$V'\left(0\right) \equiv 0$}, and $m$ is the order of the lowest
non-vanishing derivative at the origin. (In Section \ref{secso3model} we
consider a case for which the potential does not have a well-defined Taylor
expansion about the origin, but for the moment we assume that the series
above is convergent). For cases in which the origin is a maximum of the
potential, $m$ must be even, and \hbox{$d^m V / d \phi^m < 0$}. For $m$ odd,
the origin is at a saddle point, and we can define the positive $\phi$
direction to be such that \hbox{$d^m V / d \phi^m < 0$}. It is to be expected
that for most potentials arising from spontaneous symmetry breaking,
inflation will take place near an unstable maximum and $m$ will be even, but
this is an unnecessarily strict condition for the purpose of a general
analysis. The potential can be written in the form
\begin{equation}
V\left(\phi\right) = \Lambda^4 \left[1 - {1 \over m} \left({\phi \over
\mu}\right)^m + \cdots\right],\label{generalV}
\end{equation}
so that for \hbox{$\left(\phi / \mu\right) \ll 1$}, the potential is
dominated by terms of order \hbox{$\left(\phi / \mu\right)^m$}. The vacuum
energy density is \hbox{$\Lambda^4 \equiv V\left(0\right)$}, and $\mu$ is an
effective symmetry breaking scale defined by
\begin{equation}
\mu \equiv \left({(m - 1)! V\left(\phi\right) \over \left|d^m V / d
\phi^m\right|}\right)^{1 / m}\Biggr|_{\phi = 0} = v \left({(m - 1)!
f\left(x\right) \over \left|d^m f / d x^m\right|}\right)^{1 / m}\Biggr|_{x =
0},\label{effectivemassscale}
\end{equation}
We wish to evaluate the characteristics of potentials of this general form
when constrained by cosmological observations. The program is: (1) From
\hbox{$\epsilon\left(\phi_f\right) = 1$}, calculate $\phi_f$. (2) With
$\phi_f$ determined, calculate $N\left(\phi\right)$ and determine
$\phi_{60}$. (3) From $\phi_{60}$, calculate $\delta$, which constrains
$\Lambda$, and $n_s$, which constrains $\mu$. (4) From $\Lambda$, calculate
$\phi_q$ and verify \hbox{$\left(\phi_q / \phi_{60}\right) < 1$}. The cases
\hbox{$m = 2$} and \hbox{$m > 2$} exhibit strikingly different behavior, and
we consider each separately.

\subsection{Quadratic potentials}
\label{secquadraticpotentials}
Given a potential with a nonzero second derivative at the origin, \hbox{$m =
2$}, we take, for \hbox{$\left(\phi / \mu\right) \ll 1$}
\begin{equation}
V\left(\phi\right) \simeq \Lambda^4 \left[1 - {1 \over 2} \left({\phi \over
\mu}\right)^2\right],\label{quadpotential}
\end{equation}
where the effective symmetry breaking scale is
\begin{equation}
\mu \equiv \left|{V\left(0\right) \over V''\left(0\right)}\right|^{1/2}.
\end{equation}
For example, if we take a standard potential for spontaneous symmetry
breaking, \hbox{$V\left(\phi\right) = \lambda \left[\phi^2 - v^2\right]^2$},
the effective symmetry breaking scale is given by \hbox{$\mu = v / 2$}, and
the vacuum energy density is \hbox{$\Lambda^4 = \lambda v^4$}. Inflation
occurs for field values \hbox{$\phi < \phi_f$}, where
\begin{equation}
\epsilon\left(\phi_f\right) = {1 \over 16 \pi} \left(m_{Pl} \over
\mu\right)^2 \left({\left(\phi_f / \mu\right) \over 1 - \left(1/2\right)
\left(\phi_f / \mu\right)^2}\right)^2 \simeq {1 \over 16 \pi} \left(m_{Pl}
\over \mu\right)^2 \left({\phi_f \over \mu}\right)^2 \equiv 1.
\end{equation}
We then have an expression for $\phi_f$ as a function of the scale $\mu$:
\begin{equation}
\left({\phi_f \over \mu}\right) = \sqrt{16 \pi} \left({\mu \over
m_{Pl}}\right),\label{phifquad}
\end{equation}
which confirms the consistency of the approximation \hbox{$\left(\phi /
\mu\right) \ll 1$} for \hbox{$\mu \ll m_{Pl}$}. The number of e-folds
$N\left(\phi\right)$ is given by
\begin{eqnarray}
N\left(\phi\right) =&& - 8 \pi \left(\mu \over m_{Pl}\right)^2 \int_{\phi_f /
\mu}^{\phi / \mu} {{1 - x^2 / 2 \over x}dx}\cr
\simeq&& 8 \pi \left(\mu \over m_{Pl}\right)^2 \ln\left(\phi_f / \phi\right).
\end{eqnarray}
Using the value for $\phi_f$ in (\ref{phifquad}), the upper limit $\phi_{60}$
on the initial field value is
\begin{equation}
\left({\phi_{60} \over \mu}\right) = \sqrt{16 \pi} \left({\mu \over
m_{Pl}}\right) \exp\left[-{15 \over 2 \pi} \left({m_{Pl} \over
\mu}\right)^2\right],\label{phiiquad}
\end{equation}
decaying exponentially with decrease in the scale $\mu$. Scalar density
fluctuations are generated with an amplitude
\begin{eqnarray}
\delta =&& \sqrt{2 \over \pi} \left({\Lambda^2 \mu \over m_{Pl}^3}\right)
{\left[1 - \left(1 / 2\right) \left(\phi_{60} / \mu\right)^2\right]^{3/2}
\over \left(\phi_{60} / \mu\right)}\cr
\simeq&& \sqrt{2 \over \pi}\left({\mu \over m_{Pl}}\right)^3 \left({\Lambda
\over \mu}\right)^2 \left({\mu \over \phi_{60}}\right).
\end{eqnarray}
Substituting $\phi_{60}$ from (\ref{phiiquad}),
\begin{equation}
\delta = \sqrt{1 \over 8 \pi^2} \left(\mu \over m_{Pl}\right)^2 \left(\Lambda
\over \mu\right)^2 \exp\left[{15 \over 2 \pi} \left(m_{Pl} \over
\mu\right)^2\right],\label{deltaquad}
\end{equation}
with the result that the density fluctuation amplitude grows exponentially
with decreasing scale $\mu$. In order to remain consistent with the COBE DMR
result \hbox{$\delta \simeq 10^{-5}$}, we must take \hbox{$\Lambda \equiv
V\left(0\right)^{(1/4)}$} to be
\begin{eqnarray}
\left({\Lambda \over \mu}\right)^2 =&& \delta \sqrt{\pi \over 2}
\left({m_{Pl} \over \mu}\right)^3 \left({\phi_{60} \over \mu}\right)\cr
=&& \pi \sqrt{8}\,\delta \left({m_{Pl} \over \mu}\right)^2 \exp\left[-{15
\over 2 \pi} \left({m_{Pl} \over \mu}\right)^2\right].\label{Lambdaquad}
\end{eqnarray}
This illustrates the ``fine-tuning'' problem for inflationary models of this
form -- the vacuum energy density $\Lambda$ is constrained to decay
exponentially as $\left(\mu / m_{Pl}\right)$ decreases. For the potential
\hbox{$V\left(\phi\right) = \lambda \left[\phi^2 - v^2\right]^2$}, we have
\hbox{$\left(\Lambda / \mu\right)^2 = \lambda^{1/2}$}, and the constraint
(\ref{Lambdaquad}) forces the scalar coupling $\lambda$ to extremely small
values. In the sense that the parameter $\Lambda$ depends exponentially on
the symmetry breaking scale, fine-tuning is seen to be generic to potentials
of this type. The tensor fluctuation amplitude $a_T$ is:
\begin{eqnarray}
a_T =&& {2 \over \sqrt{3 \pi}} {\left[V\left(\phi_{60}\right)\right]^{1/2}
\over m_{Pl}^2}
= {2 \over \sqrt{3 \pi}} \left(\Lambda \over \mu\right)^2 \left(\mu \over
m_{Pl}\right)^2\cr
=&& 2 \delta \sqrt{8 \pi \over 3} \exp\left[- {15 \over 2 \pi} \left(m_{Pl}
\over \mu\right)^2\right],
\end{eqnarray}
and tensor fluctuations are strongly suppressed at low scale. The magnitude
of quantum fluctuations is given by:
\begin{eqnarray}
\left(\phi_q \over \mu\right) =&& \sqrt{2 \over 3 \pi} \left(\mu \over
m_{Pl}\right) \left(\Lambda \over \mu\right)^2\cr
=&& \sqrt{16 \pi \over 3} \left(m_{Pl} \over \mu\right) \exp\left[-{15 \over
2 \pi} \left(m_{Pl} \over \mu\right)^2\right].\label{phiqquad}
\end{eqnarray}
{}From (\ref{phiiquad}) we have the relationship
\begin{equation}
\left(\phi_q \over \phi_{60}\right) = {\delta \over \sqrt{3}} \left(m_{Pl}
\over \mu\right)^2,
\end{equation}
and the consistency condition \hbox{$\left(\phi_q / \phi_{60}\right) < 1$}
places a lower bound on the scale $\mu$:
\begin{equation}
\left(\mu \over m_{Pl}\right) \gtrsim \delta^{(1/2)} \simeq 3 \times 10^{-3}.
\end{equation}
A much stricter lower limit on \hbox{$\left(\mu / m_{Pl}\right)$} can,
however, be derived from the COBE limit on the scalar spectral index
\hbox{$n_s \geq 0.6$}, where, for \hbox{$\left(\phi_{60} / \mu\right) \ll
1$},
\begin{eqnarray}
n_s =&& 1 - 4 \epsilon\left(\phi_{60}\right) + 2
\eta\left(\phi_{60}\right)\cr
=&& 1 - {3 m_{Pl}^2 \over 8 \pi} \left({V'\left(\phi_{60}\right) \over
V\left(\phi_{60}\right)}\right)^2 + {m_{Pl}^2 \over 4 \pi}
\left({V''\left(\phi_{60}\right) \over V\left(\phi_{60}\right)}\right)\cr
\simeq&& 1 + {m_{Pl}^2 \over 4 \pi} \left({V''\left(\phi_{60}\right)\over
V\left(\phi_{60}\right)}\right).
\end{eqnarray}
For a potential of the form (\ref{quadpotential})
\begin{eqnarray}
1 + {m_{Pl}^2 \over 4 \pi} \left({V''\left(\phi_{60}\right)\over
V\left(\phi_{60}\right)}\right) =&& 1 - {1 \over 4 \pi} \left(m_{Pl} \over
\mu\right)^2 {1 \over 1 - \left(1 / 2\right) \left(\phi_{60} /
\mu\right)^2}\cr
\simeq&& 1 - {1 \over 4 \pi} \left(m_{Pl} \over \mu\right)^2,
\end{eqnarray}
and taking \hbox{$n_s \geq 0.6$}, we obtain the lower limit
\begin{equation}
\left(\mu \over m_{Pl}\right) \gtrsim 0.4.\label{mulimquad}
\end{equation}
This is a problematic result, however, since it precludes inflation driven by
symmetry breaking near, for instance, the grand unified scale, \hbox{$m_{GUT}
\simeq 10^{-3} m_{Pl}$}. It would be desirable to find a class of
inflationary potentials which satisfy observational constraints for scales
\hbox{$\mu \ll m_{Pl}$}. In the next section, we show that potentials of the
form (\ref{generalV}) with \hbox{$m > 2$} satisfy observational constraints
at {\it arbitrarily low} scales $\mu$, removing the substantial restriction
presented by the lower limit in equation (\ref{mulimquad}).

\subsection{Higher order potentials}
\label{sechighorderpotentials}
Now consider a potential for which the second derivative vanishes at the
origin, \hbox{$V''\left(0\right) = 0$}:
\begin{equation}
{d^n V \over d \phi^n} \biggr|_{\phi = 0} = 0\ \ (n < m),
\end{equation}
where \hbox{$m > 2$}. For small \hbox{$\left(\phi / \mu\right)$}, we can
write the potential as
\begin{equation}
V\left(\phi\right) \simeq \Lambda^4 \left[1 - {1 \over m} \left(\phi \over
\mu\right)^m\right].
\end{equation}
The effective symmetry breaking scale $\mu$ is given by equation
(\ref{effectivemassscale}). We solve for the dependence of the inflationary
constraints on the parameters $\mu$ and $\Lambda$ following the same
procedure as for the \hbox{$m = 2$} case. The first order inflationary
parameter $\epsilon$ is given by
\begin{equation}
\epsilon\left(\phi\right) = {1 \over 16 \pi} \left(m_{Pl} \over \mu\right)^2
\left({\left(\phi / \mu\right)^{(m - 1)} \over 1 - \left(1 / m\right)
\left(\phi / \mu\right)^m}\right)^2 \simeq {1 \over 16 \pi} \left(m_{Pl}
\over \mu\right)^2 \left(\phi \over \mu\right)^{2 (m - 1)}.
\end{equation}
Taking \hbox{$\epsilon\left(\phi_f\right) \equiv 1$}, we have for $\phi_f$
\begin{equation}
\left(\phi_f \over \mu\right) = \left[\sqrt{16 \pi} \left(\mu \over
m_{Pl}\right)\right]^{1 / \left(m - 1\right)}.\label{phifho}
\end{equation}
The number of e-folds $N\left(\phi\right)$ is\cite{parsons95}
\begin{eqnarray}
N\left(\phi\right) =&& - 8 \pi \left(\mu \over m_{Pl}\right)^2 \int_{\phi_f /
\mu}^{\phi / \mu} {{1 - x^m /m \over x^{m - 1}}dx}\cr
\simeq&& 8 \pi \left(\mu \over m_{Pl}\right)^2 \left(1 \over m - 2\right)
\left[\left(\mu \over \phi\right)^{m - 2} - \left(\mu \over \phi_f\right)^{m
- 2}\right].\label{Nwithfinal}
\end{eqnarray}
Substituting (\ref{phifho}) for $\phi_f$, we then have for $\phi_{60}$:
\begin{equation}
\left(\phi_{60} \over \mu\right) = \left\lbrace{15 (m - 2) \over 2 \pi}
\left(m_{Pl} \over \mu\right)^2 + \left[{1 \over \sqrt{16 \pi}} \left(m_{Pl}
\over \mu\right)\right]^{(m - 2) / (m - 1)}\right\rbrace^{-1 / (m - 2)}.
\end{equation}
Since \hbox{$\left(m - 2\right) / \left(m - 1\right) < 1$}, the
\hbox{$\left(m_{Pl} / \mu\right)^2$} term dominates, and we have the result
that $\phi_{60}$ is to a good approximation {\it independent} of $\phi_f$ for
\hbox{$\mu \ll m_{Pl}$}\cite{linde90}:
\begin{equation}
\left(\phi_{60} \over \mu\right) \simeq \left[{2 \pi \over 15 (m - 2)}
\left(\mu \over m_{Pl}\right)^2\right]^{1 / (m - 2)}.\label{phiiho}
\end{equation}
This independence will be important when we consider the consistency of the
slow-roll approximation. In this case, $\phi_{60}$ decreases as a power law
in \hbox{$\left(\mu / m_{Pl}\right)$} rather than exponentially, as in
(\ref{phiiquad}). The density fluctuation amplitude $\delta$ is
\begin{eqnarray}
\delta =&& \sqrt{2 \over \pi} \left({\Lambda^2 \mu \over m_{Pl}^3}\right)
{\left[1 - \left(1 / m\right) \left(\phi_{60} / \mu\right)^m\right]^{3/2}
\over \left(\phi_{60} / \mu\right)^{m - 1}}\cr
\simeq&& \sqrt{2 \over \pi}\left({\mu \over m_{Pl}}\right)^3 \left({\Lambda
\over \mu}\right)^2 \left({\mu \over \phi_{60}}\right)^{m - 1}.
\end{eqnarray}
Substituting $\phi_{60}$ from (\ref{phiiho}), we have
\begin{equation}
\delta = \sqrt{2 \over \pi} \left(15 (m - 2) \over 2 \pi\right)^{(m - 1) / (m
- 2)} \left(\Lambda \over \mu\right)^2 \left(\mu \over m_{Pl}\right)^{(m - 4)
/ (m - 2)}.\label{deltaho}
\end{equation}
The dependence of the density fluctuation amplitude $\delta$ on $\mu$ is then
power law, instead of expontential as in the \hbox{$m = 2$} case
(\ref{deltaquad}). For the case \hbox{$m = 4$}, which will be of interest in
the context of a specific model, the density fluctuation amplitude is
independent of \hbox{$\left(\mu / m_{Pl}\right)$}. For \hbox{$m > 4$},
$\delta$ decreases with decreasing \hbox{$\left(\mu / m_{Pl}\right)$} --
production of density fluctuations is suppressed at low scale. The COBE DMR
constraint on the vacuum energy density $\Lambda$ is then
\begin{equation}
\left({\Lambda \over \mu}\right)^2 = \delta \sqrt{\pi \over 2} \left(2 \pi
\over 15 (m - 2)\right)^{(m - 1) / (m - 2)} \left(m_{Pl} \over \mu
\right)^{(m - 4) / (m - 2)},\label{Lambdaho}
\end{equation}
and the constraint requires no fine-tuning of constants. The amplitude of
tensor fluctuations is
\begin{eqnarray}
a_T =&& {2 \over \sqrt{3 \pi}} \left(\Lambda \over \mu\right)^2 \left(\mu
\over m_{Pl}\right)^2\cr
=&& \delta \sqrt{2 \over 3} \left(2 \pi \over 15 \left(m -
2\right)\right)^{\left(m - 1\right) / \left(m - 2\right)} \left(\mu \over
m_{Pl}\right)^{m / \left(m - 2\right)},
\end{eqnarray}
and tensor fluctuations are small for \hbox{$\mu \ll m_{Pl}$}. The quantum
fluctuation amplitude $\phi_q$ is given by:
\begin{eqnarray}
\left(\phi_q \over \mu\right) =&& \sqrt{2 \over 3 \pi} \left(\mu \over
m_{Pl}\right) \left(\Lambda \over \mu\right)^2\cr
=&& {\delta \over \sqrt{3}} \left(2 \pi \over 15 (m - 2)\right)^{(m - 1) / (m
- 2)} \left(\mu \over m_{Pl}\right)^{2 / (m - 2)},
\end{eqnarray}
and the condition \hbox{$\left(\phi_q / \phi_{60}\right) < 1$} is satisfied
independent of \hbox{$\left(\mu / m_{Pl}\right)$}:
\begin{equation}
\left(\phi_q \over \phi_{60}\right) = {2 \pi \delta \over 15 \sqrt{3} (m -
2)}.
\end{equation}
The number of e-folds becomes very large at the quantum fluctuation amplitude
\begin{equation}
N\left(\phi_q\right) = 8 \pi \left(15 \over 2 \pi\right)^{m - 1} \left((m -
2) \sqrt{3} \over \delta\right)^{m - 2} \simeq 10^{5 \left(m - 2\right)}.
\end{equation}
Even for randomly selected initial conditions, as in a PNGB model, we see
that the volume of the post-inflation universe is vastly dominated by
sufficiently inflated regions, where we can take a rough upper limit on $\Pi$
in equation (\ref{probparameter}) to be
\begin{equation}
\Pi < \exp\left[180 - 3 N\left(\phi_q\right)\right] \simeq \exp\left[-10^{5
\left(m - 2\right)}\right] \ll 1.
\end{equation}
The scalar spectral index is given by:
\begin{eqnarray}
n_s \simeq&& 1 + {m_{Pl}^2 \over 4 \pi} {V''\left(\phi_{60}\right) \over
V\left(\phi_{60}\right)}\cr
\simeq&& 1 - {m - 1 \over 4 \pi} \left(m_{Pl} \over \mu\right)^2
\left(\phi_{60} \over \mu\right)^{m - 2}\cr
=&& 1 - \left(1 \over 30\right) {m - 1 \over m - 2},
\end{eqnarray}
and we have the rather surprising result that for any \hbox{$m > 2$}, the
scalar spectral index is independent of any characteristic of the potential
except the order of the lowest non-vanishing derivative at the origin. The
constraint from COBE is automatically met, with \hbox{$0.93 < n_s < 0.97$}
for all values of $m$.

One apparent difficulty with this class of potentials, however, is that the
second order slow-roll parameter $\left|\eta\right|$ becomes large for
\hbox{$\phi \ll \phi_f$}, so that the slow-roll approximation is invalid over
much of the range at which inflation is taking place. Inflation ends at the
field value $\phi_f$ given by (\ref{phifho}), but {\it slow-roll} ends at
\begin{eqnarray}
&&\left|\eta\left(\phi\right)\right| \simeq {m_{Pl}^2 \over 8 \pi}
\left|{V''\left(\phi\right) \over V\left(\phi\right)}\right| = 1\cr
&&\left(\phi \over \mu\right) = \left[{8 \pi \over m - 1} \left(\mu \over
m_{Pl}\right)^2\right]^{1 / (m - 2)} \ll \left(\phi_f \over
\mu\right).\label{endslowroll}
\end{eqnarray}
However, from equation (\ref{phiiho}), $\phi_{60}$ is {\it independent} of
$\phi_f$, so that the breakdown of slow-roll has no effect, as long as
slow-roll is valid at the initial field value,
\hbox{$\left|\eta\left(\phi_{60}\right)\right| < 1$}. If we define
$\phi_{60}$ to be 60 e-folds before the end of slow-roll as defined in
(\ref{endslowroll}), instead of the end of inflation proper, we have, using
(\ref{Nwithfinal}) for $N\left(\phi\right)$,
\begin{equation}
\left(\phi_{60} \over \mu\right) \simeq \left[{2 \pi \over 15 (m - 2)}
\left(\mu \over m_{Pl}\right)^2\right]^{1 / (m - 2)} \left(1 + {m - 1 \over
60 (m - 2)}\right)^{-{1 / (m - 2)}},
\end{equation}
which is a small correction to equation (\ref{phiiho}).

\subsection{Summary}
\label{secgeneralpotentialsummary}
In this section we summarize the results for potentials of the form
(\ref{generalV}) for the cases \hbox{$m = 2$} and \hbox{$m > 2$}. For
\hbox{$m = 2$},
\begin{eqnarray}
&&\delta = \sqrt{1 \over 8 \pi^2} \left(\mu \over m_{Pl}\right)^2
\left(\Lambda \over \mu\right)^2 \exp\left[{15 \over 2 \pi} \left(m_{Pl}
\over \mu\right)^2\right],\cr
&&n_s = 1 - {1 \over 4 \pi} \left(m_{Pl} \over \mu\right)^2,\cr
&&\left(\phi_q \over \phi_{60}\right) = {\delta \over \sqrt{3}} \left(m_{Pl}
\over \mu\right)^2\cr
&&a_T = 2 \delta \sqrt{8 \pi \over 3} \exp\left[- {15 \over 2 \pi}
\left(m_{Pl} \over \mu\right)^2\right].\label{summaryresultquad}
\end{eqnarray}
Here the COBE measurement of the scalar spectral index \hbox{$n_s \geq 0.6$}
forces the effective symmetry breaking scale to be near the Planck scale,
\hbox{$\left(\mu / m_{Pl}\right) > 0.4$}, and inflation from symmetry
breaking at low scales is inconsistent with observational constraints. For
\hbox{$m > 2$}, the corresponding result is:
\begin{eqnarray}
&&\delta = \sqrt{2 \over \pi} \left(15 (m - 2) \over 2 \pi\right)^{(m - 1) /
(m - 2)} \left(\Lambda \over \mu\right)^2 \left(\mu \over m_{Pl}\right)^{(m -
4) / (m - 2)}\cr
&&n_s = 1 - {1 \over 30} \left({m - 1 \over m - 2}\right)\cr
&&\left(\phi_q \over \phi_{60}\right) = {2 \pi \delta \over 15 \sqrt{3} (m -
2)}\cr
&&a_T = \delta \sqrt{2 \over 3} \left(2 \pi \over 15 \left(m -
2\right)\right)^{\left(m - 1\right) / \left(m - 2\right)} \left(\mu \over
m_{Pl}\right)^{m / \left(m - 2\right)}.\label{summaryresultho}
\end{eqnarray}
The scalar spectral index $n_s$ is independent of any characteristic of the
potential except $m$. For the case \hbox{$m = 4$}, the density fluctuation
amplitude is independent of \hbox{$\left(\mu / m_{Pl}\right)$}, and inflation
can take place successfully at an arbitrary symmetry breaking scale. In
Section \ref{secso3model}, we illustrate this behavior within the context of
a specific model, in which the potential is created by loop effects in a
Lagrangian with an explicitly broken SO(3) symmetry.

It should be noted that it is not strictly necessary that $V''(0)$ vanish for
inflation to be characterized by (\ref{summaryresultho}) for some range of
effective symmetry breaking scales $\mu$. It is sufficient that the second
derivative at the origin be small relative to some derivative of order
\hbox{$m > 2$}. If there is a range of $\phi_{60}$ such that
\begin{equation}
{1 \over 2} \left|{V''\left(0\right) \over V(0)}\right| \phi_{60}^2 \ll {1
\over m!} {1 \over V(0)} \left|{d^m V \over d \phi^m}\right|_{\phi = 0}
\phi_{60}^m \equiv {1 \over m} \left(\phi_{60} \over \mu\right)^m,
\end{equation}
then the potential is still be dominated by terms of order $\phi^m$ for
\hbox{$\phi \geq \phi_{60}$}, and $V\left(\phi\right)$ is of the form
(\ref{generalV}) to a good approximation for field values of physical
interest. Taking $\phi_{60}$ to be approximately of the form (\ref{phiiho}),
we have
\begin{equation}
{2 \pi \over 15 \left(m - 2\right)} \left(\mu \over m_{Pl}\right)^2 > \mu^2
\left(m \over 2\right) \left|{V''(0) \over V(0)}\right| = 4 \pi m \left(\mu
\over m_{Pl}\right)^2 \left|\eta\left(0\right)\right|.
\end{equation}
We then have a constraint on the value of the second order slow-roll
parameter $\eta$ at the origin
\begin{equation}
\left|\eta\left(0\right)\right| < {1 \over 30 m \left(m -
2\right)}.\label{socondition}
\end{equation}
If we write $V\left(\phi\right)$ in terms of a dimensionless function $f$ as
in (\ref{dimensionlessf}), the parameter $\eta\left(0\right)$ is given by
\begin{equation}
\eta\left(0\right) = {1 \over 8 \pi} \left(m_{Pl} \over v\right)^2
{f''\left(0\right) \over f\left(0\right)} \propto \left(m_{Pl} \over
v\right)^2,
\end{equation}
resulting in a lower limit on the symmetry breaking scale $v$,
\begin{equation}
\left(v \over m_{Pl}\right)^2 < {15 m \left(m - 2\right) \over 4 \pi} \left|
{f''\left(0\right) \over f\left(0\right)} \right|\label{mixedlimit},
\end{equation}
which depends on the particular form of the potential. This is illustrated in
the context of a specific model in Section \ref{secso3model}.

\section{Inflation from SO(3) pseudo Nambu-Goldstone bosons}
\label{secso3model}
It is not immediately clear that potentials with \hbox{$m > 2$} can be
generated by spontaneous symmetry breaking, since such symmetry breaking is,
at least at tree level, created by scalar mass terms. A Lagrangian of the
generic form
\begin{equation}
{\cal L} = \left(\partial_\mu \phi\right)^{\dag} \left(\partial^\mu
\phi\right) - {1 \over 2} \mu^2 \phi^2 - {\lambda \over 4} \phi^4
\end{equation}
only exhibits spontaneous symmetry breaking at tree level for \hbox{$\mu^2 <
0$}, and the potential is automatically dominated by quadratic terms near
\hbox{$\phi = 0$}. One physically well motivated possibility for overcoming
this difficulty is a model involving scalar particles which are massless at
tree level, but which acquire mass through radiative corrections. The
original models for new inflation\cite{linde82,albrect82} are of this type,
using Coleman-Weinberg symmetry breaking to create the inflationary
potential, with \hbox{$m = 4$}. Natural inflation models, using pseudo
Nambu-Goldstone bosons to drive inflation, also belong to this category. Here
we consider a natural inflation model in which the PNGB potential is created
by loop effects from gauge bosons.

\subsection{Pseudo Nambu-Goldstone bosons from gauge boson loops}
\label{secPNGBs}
Take a scalar particle Lagrangian which is invariant under some spontaneously
broken gauge group G:
\begin{eqnarray}
&&{\cal L} = \left(D_\mu \phi\right)^{\dag} \left(D^\mu \phi\right) - \lambda
\left[\phi^{\dag} \phi - v^2\right]^2 - {1 \over 4} Tr\left[F_{\mu \nu}
F^{\mu \nu}\right]\cr
&&D_\mu \equiv \partial_\mu - i g Q_j A^j_\mu\cr
&&F^i_{\mu \nu} \equiv \partial_\mu A^i_\nu - \partial_\nu A^i_\mu + g C_{i j
k} A^j_\mu A^k_\nu,\label{genericgaugeL}
\end{eqnarray}
where G has generators \hbox{$\left\lbrace Q_i \right\rbrace$} with
commutation relation \hbox{$\left[Q_i, Q_j\right] = i C_{i j k} Q_k$}. We
take $\phi$ in a vector representation \hbox{$\phi \equiv \left(\phi_1,
\ldots \phi_n\right)$}, which transforms under ${\rm G}$ as
\begin{equation}
\phi \rightarrow \exp\left[i Q_k \xi^k\right] \phi.
\end{equation}
The product \hbox{$\phi^{\dag} \phi$} is then manifestly invariant under the
group G. There is one gauge boson $A^i_\mu$ for each generator of the group
G, with the gauge transformation law
\begin{equation}
A_\mu^i \rightarrow A_\mu^i+ {1 \over g} \partial_\mu \xi^i +  C_{i j k}
A_\mu^j \xi^k.
\end{equation}
For the Lagrangian (\ref{genericgaugeL}), the group symmetry G is
spontaneously broken when $\phi$ acquires a nonzero vacuum expectation value,
\hbox{$<\phi> = v$}. We can parameterize $\phi$ in terms of a ``shifted''
field \hbox{$<\sigma> = (0,\ldots,0)$} and massless Nambu-Goldstone bosons
$\xi^i$ ({\it unitary gauge}):
\begin{equation}
\phi = \exp\left[i Q_k \xi^k\right] \left[\sigma + {\bf v}\right],
\end{equation}
where ${\bf v}$ is the vacuum expectation value \hbox{${\bf v} \equiv
(v_1,\ldots,v_n),\,{\bf v}^{\dag} {\bf v} = v^2$}. In the spontaneously
broken vacuum, the gauge bosons acquire a mass
\begin{equation}
M^2_{i j} = g^2 {\bf v}^{\dag} Q_i Q_j {\bf v},
\end{equation}
and the Nambu-Goldstone modes $\xi^i$ which correspond to the broken
generators of G are ``eaten'' to form longitudinal modes for the gauge bosons
$A_\mu$.

The gauge group, however, does not necessarily need to be the full symmetry
group G. It is consistent to have gauge bosons which transform under some
{\it subgroup} \hbox{${\rm {\bar G}} \subset {\rm G}$}, with generators
\hbox{$\left\lbrace {\bar Q_i}\right\rbrace \subset \left\lbrace
Q_i\right\rbrace$}. If we take
\begin{equation}
D_\mu \equiv \partial_\mu - i g {\bar Q_j} A^j_\mu,
\end{equation}
where \hbox{$j = 1,\ldots,dim\left({\rm {\bar G}}\right)$}, there are in
general as many as \hbox{$dim\left({\rm G}\right) - dim\left({\rm {\bar
G}}\right)$} leftover massless Nambu-Goldstone bosons $\xi$. However, since
the symmetry ${\rm G}$ of the scalar potential is not a symmetry of the
entire Lagrangian, the leftover modes are PNGB's, and do not in general
remain massless when radiative corrections are taken into account. This is
reflected in the fact that the gauge boson mass matrix depends on the PNGB
fields
\begin{equation}
M^2_{i j}\left(\xi\right) = g^2 {\bf v}^{\dag} \exp\left[-i Q_k \xi^k\right]
{\bar Q_i} {\bar Q_j} \exp\left[i Q_k \xi^k\right] {\bf v},
\end{equation}
where \hbox{$i,j = 1,\ldots,dim\left({\rm {\bar G}}\right)$}, and \hbox{$k >
dim\left({\rm {\bar G}}\right)$}. Gauge boson loop effects generate a
one-loop effective potential of the form\cite{weinberg73}
\begin{equation}
V_1\left(\xi\right) = {3 \over 64 \pi^2} Tr\left\lbrace
\left[M^2\left(\xi\right)\right]^2 \ln\left[M^2\left(\xi\right) \over
v^2\right]\right\rbrace.
\end{equation}
We use an effective potential of this type to drive inflation. Note that the
dependence of $M^2_{i j}$ on the modes $\xi^k$ disappears if the commutator
\hbox{$\left[Q_k, {\bar Q_j}\right]$} vanishes for all \hbox{$j =
1,\ldots,dim\left({\rm {\bar G}}\right),\, k = dim\left({\rm {\bar G}}\right)
+ 1,\ldots,dim\left({\rm G}\right)$}, so that the gauge group ${\rm {\bar
G}}$ must be a nontrivial subgroup of the scalar symmetry group ${\rm G}$. In
particular, if ${\rm G}$ is a direct product group, \hbox{${\rm G} = {\rm
G}_1 \bigotimes {\rm G}_2$}, and \hbox{${\bar{\rm G}} = {\rm G}_1$}, the
dependence of $V_1$ on $\xi$ vanishes.

\subsection{PNGB's from an explicitly broken SO(3) symmetry}
\label{secSO3PNGBs}
Take a Lagrangian with three real scalar fields $\phi_1, \phi_2, \phi_3$ and
an SO(3) symmetric potential
\begin{equation}
V\left(\phi\right) \equiv \lambda \left[\phi_1^2 + \phi_2^2 + \phi_3^2 -
v^2\right]^2.\label{scalarpotentiala}
\end{equation}
It is convenient to parameterize the fields as a triplet
\begin{equation}
\phi \equiv \left(\matrix{\phi^+\cr\phi^0\cr\phi-}\right),
\end{equation}
where
\begin{eqnarray}
\phi^\pm \equiv&& {1 \over \sqrt{2}} \left(\phi_1 \pm \phi_2\right),\cr
\phi^0 \equiv&& \phi_3.
\end{eqnarray}
In this basis, the generators of the SO(3) symmetry group can be taken to be
\begin{eqnarray}
&&T_1 = {1 \over \sqrt{2}} \left(\matrix{\ 0&-1&\ 0\cr -1&\ 0&\ 1\cr \ 0&\
1&\ 0}\right),\cr
&&T_2 = {1 \over \sqrt{2}} \left(\matrix{\ 0&-i&\ 0\cr \ i&\ 0&\ i\cr \
0&-i&\ 0}\right),\cr
&&T_3 = \left(\matrix{\ 1&\ 0&\ 0\cr \ 0&\ 0&\ 0\cr \ 0&\ 0&-1}\right),
\end{eqnarray}
with commutation relation
\begin{equation}
\left[T_i, T_j\right] = -i \epsilon_{i j k} T_k,
\end{equation}
where $\epsilon_{i j k}$ is the Levi-Civita tensor. Note that $T_3$ generates
a ${\rm U(1)}$ subgroup of ${\rm SO(3)}$, where the $\phi^\pm$ fields are
charged under the ${\rm U(1)}$ and the $\phi^0$ field is neutral. We take the
${\rm U(1)}$ generated by $T_3$ to be the gauge group, with a Lagrangian of
the form
\begin{eqnarray}
&&{\cal L} = \left(D_\mu \phi\right)^{\dag}\left(D^\mu \phi\right) - \lambda
\left[\phi^{\dag} \phi - v^2\right]^2 - {1 \over 4} F_{\mu \nu} F^{\mu
\nu},\cr
&&D_\mu \equiv \partial_\mu - i g T_3 A_\mu,\cr
&&F_{\mu \nu} \equiv \partial_\mu A_\nu - \partial_\nu A_\mu.
\end{eqnarray}
Here we a have a {\it single} gauge boson $A_\mu$, transforming under the
${\rm U(1)}$ gauge symmetry generated by $T_3$, which explicitly breaks the
${\rm SO(3)}$ symmetry of the scalar potential. We choose the following
general parameterization for $\phi$:
\begin{eqnarray}
\phi \equiv&& \left(\matrix{\phi^+\cr\phi^0\cr\phi-}\right) = \left[\sigma +
v\right] \left(\matrix{(1 / \sqrt{2}) e^{i \left(\alpha / v\right)}
\sin\left(\theta / v\right)\cr \cos\left(\theta / v\right)\cr (1 / \sqrt{2})
e^{-i \left(\alpha / v\right)} \sin\left(\theta / v\right)}\right)\cr
=&& e^{i T_3 \left(\alpha / v\right)} \left[\sigma + v\right]
\left(\matrix{(1 / \sqrt{2}) \sin\left(\theta / v\right)\cr \cos\left(\theta
/ v\right)\cr (1 / \sqrt{2}) \sin\left(\theta /
v\right)}\right),\label{vacparam}
\end{eqnarray}
which can be recognized as spherical coordinates in the basis
\hbox{$\left(\phi_1, \phi_2, \phi_3\right)$}. The modes $\alpha$ and $\theta$
are the Nambu-Goldstone bosons. Note that although the ${\rm SO(3)}$ symmetry
has three generators, there are only two Nambu-Goldstone bosons, since the
${\rm SO(3)}$ is spontaneously broken to ${\rm SO(2)}$, which has one
generator. This residual symmetry corresponds to rotating the vacuum
expectation vector ${\bf v}$ about itself, where
\begin{equation}
{\bf v} \equiv v e^{i T_3 \left(\alpha / v\right)} \left(\matrix{(1 /
\sqrt{2}) \sin\left(\theta / v\right)\cr \cos\left(\theta / v\right)\cr (1 /
\sqrt{2}) \sin\left(\theta / v\right)}\right).\label{vacexpec}
\end{equation}
Note that the residual ${\rm SO(2)}$ symmetry of the vacuum {\it cannot} in
general be identified with the ${\rm U(1)}$ gauge symmetry. In the
spontaneously broken phase, the $\alpha$ mode is absorbed by the gauge boson
$A_\mu$, which acquires a mass
\begin{eqnarray}
M^2 =&& g^2 \left({\bf v}^{\dag} T_3^2 {\bf v}\right)\cr
=&& g^2 v^2 \sin^2\left(\theta \over v\right).\label{so3gaugemass}
\end{eqnarray}
Loop effects generate a one-loop effective potential
\begin{eqnarray}
V_1\left(\theta\right) \equiv&& {3 \over 64 \pi^2}
\left[M^2\left(\theta\right)\right]^2 \ln\left[M^2\left(\theta\right) \over
v^2\right]\cr
=&& {3 v^4 \over 64 \pi^2} g^4 \sin^4\left(\theta \over v\right) \ln\left[g^2
\sin^2\left(\theta \over v\right)\right],
\end{eqnarray}
which has a maximum at \hbox{$\theta = 0$} and a minimum at \hbox{$\theta =
\pi v / 2$}, which is the physical vacuum. However, for a perturbative
coupling, \hbox{$g < 1$}, the potential is negative at the physical vacuum,
so we normalize to adjust the vacuum energy at the minimum to zero:
\begin{eqnarray}
V\left(\theta\right) \equiv&& V_1\left(\theta\right) - V_1\left(\pi v /
2\right)\cr
=&& {3 v^4 \over 64 \pi^2} g^4 \left\lbrace \sin^4\left(\theta \over v\right)
\ln\left[g^2 \sin^2\left(\theta \over v\right)\right] -
\ln\left(g^2\right)\right\rbrace\label{so3pot}
\end{eqnarray}
We take this to be the inflationary potential, neglecting any effects due to
contributions from a fermionic sector, which will be discussed later.

\subsection{Inflation from SO(3) gauge potentials}
\label{secso3inflation}
For \hbox{$\left(\theta / v\right) \ll 1$}, the PNGB potential (\ref{so3pot})
becomes
\begin{equation}
V\left(\theta\right) \simeq {3 v^4 \over 64 \pi^2} g^4 \left\lbrace
\left(\theta \over v\right)^4 \ln\left[g^2 \left(\theta \over
v\right)^2\right] - \ln\left(g^2\right)\right\rbrace,\label{so3approxpot}
\end{equation}
which is dominated by terms of order \hbox{$\left(\theta / v\right)^4$}, so
we expect the inflationary constraints to be described by equation
(\ref{summaryresultho}). However, the potential (\ref{so3approxpot}) does
{\it not} have a well-defined Taylor expansion about the origin, since the
fourth derivative, describing the PNGB self coupling, diverges
logarithmically as \hbox{$\theta \rightarrow 0$}:
\begin{eqnarray}
{d^m V \over d \theta^m} \biggr|_{\theta \rightarrow 0} =&& 0\ \ (m \leq
3),\cr
{d^4 V \over d \theta^4} \biggr|_{\theta \rightarrow 0} \propto&&
\ln\left(\theta \over v\right) \rightarrow -\infty.
\end{eqnarray}
The origin of this behavior is the familiar infared divergence in the gauge
boson propagator, since the ${\rm U(1)}$ gauge symmetry is unbroken at
\hbox{$\theta = 0$} and the gauge boson $A_\mu$ is massless. However,
$V\left(\theta\right)$ does have a well-defined Taylor expansion about any
{\it finite} field value $\theta_0$, for which the gauge symmetry is broken
and $A_\mu$ acquires a mass. For a field value \hbox{$\theta \simeq
\theta_0$}, the potential is of the form
\begin{equation}
V\left(\theta\right) \simeq {3 v^4 \over 64 \pi^2} g^4 \left[\left(\theta
\over \mu\right)^4 - \ln\left(g^2\right)\right],
\end{equation}
where $\mu$ is a redefined mass scale which depends on the choice of
$\theta_0$. An appropriate value of $\theta_0$ can be chosen as a function of
the symmetry breaking scale $v$, and the results (\ref{summaryresultho}) are
valid up to logarithmic corrections. The corrections can be determined
iteratively.
The potential can be written in the form:
\begin{equation}
V\left(\theta\right) = - {3 v^4 \over 64 \pi^2} g^4 \ln\left(g^2\right)
\left[1 - {1 \over 4} \left(\theta \over \mu_0\right)^4\right] + {3 v^4 \over
256 \pi^2} g^4 \left(\theta \over \mu_0\right)^4 \ln\left[{1 \over 2}
\left(\theta \over \mu_0\right)^2\right],
\end{equation}
where \hbox{$\mu_0 \equiv v / \sqrt{2}$}. Expanding the logarithm about a
field value \hbox{$\theta = \theta_0$},
\begin{eqnarray}
\ln\left[{1 \over 2} \left(\theta \over \mu_0\right)^2\right] =&& \ln\left[{1
\over 2} \left(\theta_0 \over \mu_0\right)^2\right] - 2 \sum_{n = 1}^\infty
{1 \over n} \left(1 - {\theta \over \theta_0}\right)^n\cr
\simeq&& \ln\left[{1 \over 2} \left(\theta_0 \over \mu_0\right)^2\right]\ \
\left(\theta \simeq \theta_0\right).\label{logexpansion}
\end{eqnarray}
For \hbox{$\theta \simeq \theta_0$}, the potential is then of the form
(\ref{generalV})
\begin{eqnarray}
V\left(\theta\right) \simeq&& - {3 v^4 \over 64 \pi^2} g^4
\ln\left(g^2\right) \left[ 1 - {1 \over 4} \left(\theta \over \mu_0\right)^4
\left(1 + {\ln\left[\left(1/2\right) \left(\theta_0 / \mu_0\right)^2\right]
\over \ln\left(g^2\right)}\right)\right]\cr
=&& \Lambda^4 \left[1 - {1 \over 4} \left(\theta \over \mu_1\right)^4\right],
\end{eqnarray}
where the vacuum energy density is
\begin{equation}
\Lambda^4 \equiv - {3 v^4 \over 64 \pi^2} g^4 \ln\left(g^2\right),
\end{equation}
and the scale $\mu_1$ is
\begin{equation}
\mu_1 \equiv \mu_0 \left(1 + {\ln\left[\left(1/2\right) \left(\theta_0 /
\mu_0\right)^2\right]  \over \ln\left(g^2\right)}\right)^{-1/4}.
\end{equation}
The scale $\mu_1$ depends on the expansion parameter $\theta_0$ and the gauge
coupling g. Since all the quantities of physical interest are defined in
terms of the field value \hbox{$\theta = \theta_{60}$}, we take the expansion
parameter $\theta_0$ to be the value of $\theta_{60}$ from equation
(\ref{phiiho}) with m = 4:
\begin{equation}
\left(\theta_0 \over \mu_0\right) = \sqrt{\pi \over 15} \left(\mu_0 \over
m_{Pl}\right).\label{theta0}
\end{equation}
Similarly, we determine the coupling constant to lowest order  \hbox{$g
\simeq g_0$} from (\ref{Lambdaho}):
\begin{equation}
\left(\Lambda \over \mu_0\right)^4 \equiv - {3 \over 16 \pi^2} g_0^4
\ln\left(g_0^2\right) = {\pi \over 2} \left(\pi \over 15\right)^3 \delta^2,
\end{equation}
which is conveniently independent of $\mu_0$. Note that the limit on
$\Lambda$ from the COBE value \hbox{$\delta \simeq 10^{-5}$} serves in this
context to constrain the gauge coupling $g$. To lowest order, the gauge
coupling is independent of the symmetry breaking scale $v$, where
\begin{equation}
g_0^4 \ln\left(g_0^2\right) = - {8 \over 3} \left(\pi^2 \over 15\right)^3
\delta^2,\label{gloworder}
\end{equation}
and we take the first order scale $\mu_1$ to be
\begin{equation}
\mu_1 = \mu_0 \left(1 + {\ln\left[\left(\pi / 30\right) \left(\mu_0 /
m_{Pl}\right)^2\right]  \over
\ln\left(g_0^2\right)}\right)^{-1/4}.\label{mufirstorder}
\end{equation}
We can then use (\ref{phiiho}) to obtain the first order correction to
$\theta_{60}$:
\begin{eqnarray}
\left(\theta_{60} \over \mu_1\right) =&& \sqrt{\pi \over 15} \left(\mu_1
\over m_{Pl}\right)\cr
\Longrightarrow \left(\theta_{60} \over \mu_0\right) =&& \sqrt{\pi \over 15}
\left(\mu_0 \over m_{Pl}\right) \left(\mu_1 \over \mu_0\right)^2.
\end{eqnarray}
{}From (\ref{mufirstorder}), substituting \hbox{$\mu_0 \equiv v / \sqrt{2}$},
\begin{equation}
\left(\theta_{60} \over v\right) = {1 \over 2} \sqrt{\pi \over 15} \left(v
\over m_{Pl}\right) \left(1 + {\ln\left[\left(\pi / 60\right) \left(v /
m_{Pl}\right)^2\right]  \over
\ln\left(g_0^2\right)}\right)^{-1/2},\label{thetaifirstorder}
\end{equation}
where $g_0$ satisfies equation (\ref{gloworder}). We can also obtain a first
order correction to the gauge coupling by using
\begin{eqnarray}
\left(\Lambda \over \mu_1\right)^4 = \left(\mu_0 \over \mu_1\right)^4
\left(\Lambda \over \mu_0\right)^4 =&& - {3 \over 16 \pi^2} \left(\mu_0 \over
\mu_1\right)^4 g^4 \ln\left(g^2\right)\cr
=&& {\pi \over 2} \left(\pi \over 15\right)^3 \delta^2,
\end{eqnarray}
so that to first order, \hbox{$g \simeq g_1$} satisfies
\begin{eqnarray}
g_1^4 \ln\left(g_1^2\right) =&& - {8 \over 3} \left(\pi^2 \over 15\right)^3
\delta^2 \left(\mu_1 \over \mu_0\right)^4\cr
=&& - {8 \over 3} \left(\pi^2 \over 15\right)^3 \delta^2 \left(1 +
{\ln\left[\left(\pi / 60\right) \left(v / m_{Pl}\right)^2\right]  \over
\ln\left(g_0^2\right)}\right)^{-1},
\end{eqnarray}
and the COBE limited coupling constant does not require fine-tuning, with
\hbox{$g \simeq 10^{-3}$} for a wide range of symmetry breaking scales $v$.
Since \hbox{$\left(\theta_{60} / \mu_1\right) \propto \left(\mu_1 /
m_{Pl}\right)$}, the scalar spectral index is insensitive to the redefinition
of scale, remaining independent of $v$:
\begin{eqnarray}
n_s =&& 1 - {3 \over 4 \pi} \left(m_{Pl} \over \mu_1\right)^2
\left(\theta_{60} \over \mu_1\right)^2 = 1 - {1 \over 20}\cr
=&& 0.95.
\end{eqnarray}
The expansion (\ref{logexpansion}) is valid only if the series
\begin{equation}
\sum_{n = 1}^\infty {1 \over n} \left(1 - {\theta \over \theta_0}\right)^n
\end{equation}
converges at \hbox{$\theta = \theta_{60}$} from (\ref{thetaifirstorder}).
Taking $\theta_0$ from (\ref{theta0}),
\begin{equation}
\left(\theta_{60} \over \theta_0\right) = \left(1 + {\ln\left[\left(\pi /
60\right) \left(v / m_{Pl}\right)^2\right] \over
\ln\left(g_0^2\right)}\right)^{-1/2},
\end{equation}
which is less than one for all \hbox{$v < m_{Pl}$}, and (\ref{logexpansion})
is consistent.

We then have the result that with a logarithmic divergence near the origin,
the result (\ref{summaryresultho}) is approximately valid, with corrections
derivable by an iterative solution:
\begin{eqnarray}
&&\left(\theta_{60} \over v\right) = {1 \over 2} \sqrt{\pi \over 15} \left(v
\over m_{Pl}\right) \left(1 + {\ln\left[\left(\pi / 60\right) \left(v /
m_{Pl}\right)^2\right]  \over \ln\left(g_0^2\right)}\right)^{-1/2}\cr
&&g^4 \ln\left(g^2\right) = g_0^4 \ln\left(g_0^2\right) \left(1 +
{\ln\left[\left(\pi / 60\right) \left(v / m_{Pl}\right)^2\right]  \over
\ln\left(g_0^2\right)}\right)^{-1}\cr
&&g_0^4 \ln\left(g_0^2\right) \equiv - {8 \over 3} \left(\pi^2 \over
15\right)^3 \delta^2\cr
&&n_s = 0.95
\end{eqnarray}
Of particular note, the scalar spectral index $n_s$ remains independent of
the symmetry breaking scale $v$.

\subsection{Inclusion of fermions}
\label{secfermions}
In this section we discuss the coupling of fermionic species to the scalars
$\phi$, and discuss the effects on the inflationary constraints which we
obtained by considering only the effective potential generated by gauge boson
loops. Clearly, we can preserve all of the previous results exactly if the
Fermi sector is invariant under the full ${\rm SO(3)}$ symmetry group:
\begin{eqnarray}
{\cal L}_F =&& i {\bar\psi} \gamma^\mu D_\mu \psi - h \sum_{n = 1}^3
\left({\bar\psi}_L \phi \psi_{n R} + {\bar\psi}_{n R} \phi^{\dag}
\psi_L\right),\cr
D_\mu \equiv&& \partial_\mu - i g T_3 A_\mu,
\end{eqnarray}
where the $\psi_{n R}$ are singlet fields and $\psi_L$ is a triplet
\begin{equation}
\psi_L \equiv \left(\matrix{\psi_{1 L}\cr\psi_{2 L}\cr\psi_{3 L}}\right),
\end{equation}
which transforms under the ${\rm SO(3)}$ as
\begin{equation}
\psi_L \rightarrow \exp\left[i T_k \xi^k\right] \psi_L.
\end{equation}
In this case the fermion mass is independent of the value of the PNGB mode
$\theta$, and there is no correction to the effective potential
(\ref{so3pot}) except for the addition of a constant.

However, we can couple fermions which also explicitly break the ${\rm SO(3)}$
symmetry, as long as the Lagrangian respects the ${\rm U(1)}$ gauge symmetry.
The simplest fermion Lagrangian of this type is of the form
\begin{equation}
{\cal L}_F = i {\bar\psi}_1 \gamma^\mu D_\mu \psi_1 + i {\bar\psi}_2
\gamma^\mu \partial_\mu \psi_2 - h_1 \left({\bar\psi}_{1L} \phi^+ \psi_{1R} +
{\bar\psi}_{1R} \phi^- \psi_{1L}\right) - h_2 {\bar\psi}_2 \phi^0 \psi_2,
\end{equation}
where $\psi_1$ is charged under the ${\rm U(1)}$ gauge group and $\psi_2$ is
neutral:
\begin{eqnarray}
\psi_{1L} \rightarrow&& e^{i \alpha / 2} \psi_{1L}\ \ \psi_{1R} \rightarrow
e^{-i \alpha / 2} \psi_{1R},\cr
\psi_2 \rightarrow&& \psi_2.
\end{eqnarray}
In the spontaneously broken phase, where
\begin{eqnarray}
\phi^\pm =&& {v \over \sqrt{2}} \sin\left(\theta \over v\right),\cr
\phi^0 =&& v \cos\left(\theta \over v\right),
\end{eqnarray}
the fermions acquire masses
\begin{eqnarray}
m_1^2 = {h_1^2 v^2 \over 2} \sin^2\left(\theta \over v\right),\cr
m_2^2 = h_2^2 v^2 \cos^2\left(\theta \over v\right).
\end{eqnarray}
The one-loop effective potential generated by fermion loops is then
\begin{eqnarray}
V_{1F} = &&- {1 \over 16 \pi^2} m_1^4 \ln\left(m_1^2 \over v^2\right) - {1
\over 16 \pi^2} m_2^4 \ln\left(m_2^2 \over v^2\right)\cr
= &&- {1 \over 64 \pi^2} v^4 h_1^4 \sin^4\left(\theta \over v\right)
\ln\left[{h_1^2 \over 2} \sin^2\left(\theta \over v\right)\right]\cr
&&- {1 \over 16 \pi^2} v^4 h_2^4 \cos^4\left(\theta \over v\right)
\ln\left[h_2^2 \cos^2\left(\theta \over v\right)\right].
\end{eqnarray}
Note that the potential generated by the charged fermions $\psi_1$ is of the
same form as the one-loop gauge potential (\ref{so3pot}), and we can write
the full effective potential, including gauge boson loop contributions, in
the form
\begin{eqnarray}
V\left(\theta\right) = &&{v^4 \over 64 \pi^2} \left(3 g^4 - h_1^4\right)
\sin^4\left(\theta \over v\right) \ln\left[\sin^2\left(\theta \over
v\right)\right]\cr
&&+{v^4 \over 64 \pi^2} \left[3 g^4 \ln\left(g^2\right) - h_1^4
\ln\left({h_1^2 \over 2}\right)\right] \sin^4\left(\theta \over v\right)\cr
&&-{v^4 \over 16 \pi^2} h_2^4 \cos^4\left(\theta \over v\right)
\ln\left[h_2^2 \cos^2\left(\theta \over v\right)\right]\cr
&&-{v^4 \over 64 \pi^2} \left[3 g^4 \ln\left(g^2\right) - h_1^4
\ln\left({h_1^2 \over 2}\right)\right].
\end{eqnarray}
The only effect of the coupling of $\phi$ to the charged fermion $\psi_1$ is
to add a correction to the vacuum energy density \hbox{$\Lambda^4 \equiv
V\left(0\right)$}:
\begin{equation}
\Lambda^4 = - {v^4 \over 64 \pi^2} \left[3 g^4 \ln\left(g^2\right) - h_1^4
\ln\left({h_1^2 \over 2}\right) + 4 h_2^4 \ln\left(h_2^2\right)\right].
\end{equation}
For \hbox{$h_1 \gtrsim g$}, the potential changes sign, and the minimum is at
\hbox{$\theta = 0$}. However, we exclude this parameter region from
consideration because the mass of the gauge boson $A_\mu$
(\ref{so3gaugemass}) vanishes at the origin, and the resulting theory
contains unacceptable long-range forces. If we take \hbox{$h_1 \ll g$}, the
correction to the potential due to the fermion $\psi_1$ is negligible, and we
can write the potential as
\begin{eqnarray}
V\left(\theta\right) = &&{3 v^4 \over 64 \pi^2} \left\lbrace g^4
\sin^4\left(\theta \over v\right) \ln\left[g^2 \sin^2\left(\theta \over
v\right)\right] - g^4 \ln\left(g^2\right)\right\rbrace\cr
&&- {v^4 \over 16 \pi^2} h^4 \cos^4\left(\theta \over v\right) \ln\left[h^2
\cos^2\left(\theta \over v\right)\right],
\end{eqnarray}
where \hbox{$h \equiv h_2$}. For \hbox{$\left(\theta / v\right) \ll 1$}, the
coupling to the neutral fermion $\psi_2$ introduces quadratic terms into the
potential:
\begin{eqnarray}
V\left(\theta\right) \simeq &&{v^4 \over 64 \pi^2} \left[ 3 g^4
\ln\left(g^2\right) \left(\theta \over v\right)^4 + 8 h^4 \ln\left(h^2\right)
\left(\theta \over v\right)^2\right]\cr
&&+ {v^4 \over 64 \pi^2} \left\lbrace 3 g^4 \left(\theta \over v\right)^4
\ln\left[\left(\theta \over v\right)^2\right]\right\rbrace\cr
&&-{v^4 \over 64 \pi^2} \left[3 g^4 \ln\left(g^2\right) + 4 h^4
\ln\left(h^2\right)\right]\ \ \left(\theta \ll v\right).
\end{eqnarray}
However, for \hbox{$h \ll g$}, there is still a range of symmetry breaking
scales for which the potential is dominated by terms of order
\hbox{$\left(\theta / v\right)^4$} at \hbox{$\theta = \theta_{60}$}:
\begin{equation}
3 g^4 \ln\left(g^2\right) \left(\theta_{60} \over v\right)^4 > 8 h^4
\ln\left(h^2\right) \left(\theta_{60} \over v\right)^2.
\end{equation}
Using the lowest order result for $\theta_{60}$,
\begin{equation}
\left(\theta_{60} \over v\right) = {1 \over 2} \sqrt{\pi \over 15} \left(v
\over m_{Pl}\right),
\end{equation}
we obtain a lower limit on the symmetry breaking scale
\begin{equation}
\left(v \over m_{Pl}\right) > \left(h \over g\right)^2 \sqrt{160
\ln\left(h^2\right) \over \pi \ln\left(g^2\right)},
\end{equation}
which is just the condition (\ref{mixedlimit}). For \hbox{$g \simeq
10^{-3}$}, taking \hbox{$h \simeq 10^{-6}$} gives a lower limit \hbox{$v
\gtrsim 10^{-5} m_{Pl} \simeq 10^{14} {\rm GeV}$}, low enough for symmetry
breaking at the grand unified scale, \hbox{$m_{GUT} \simeq 10^{16} {\rm
GeV}$}.

\section{Conclusions}

For scalar field potentials $V\left(\phi\right)$ which possess an unstable
equilibrium at the origin and a minimum characterized by a symmetry breaking
scale $v$, we have shown that for \hbox{$v \ll m_{Pl}$}, the entire period of
inflation is characterized by \hbox{$\phi \ll v$}, and the potential can be
expressed in the form (\ref{generalV}).
For the case \hbox{$m = 2$}, the observable quantities produced during
inflation are then given by equations (\ref{summaryresultquad}).
The COBE limit on the scalar spectral index, \hbox{$n_s \geq 0.6$}, places a
lower limit on the effective symmetry breaking scale \hbox{$\left(\mu /
m_{Pl}\right) > 0.4$}. In addition, the density fluctuation amplitude
$\delta$  depends exponentially on \hbox{$\left(\mu / m_{Pl}\right)$}, and
fine-tuning of the parameter $\Lambda$ is required to suppress production of
density fluctuations to the level of the COBE value \hbox{$\delta \simeq
10^{-5}$}.
For potentials with a vanishing second derivative at the origin, \hbox{$m >
2$}, the corresponding result is given by equations (\ref{summaryresultho}).
In this case, he scalar spectral index $n_s$ is {\it independent} of any
characteristic of the potential except the order of the lowest non-vanishing
derivative, and is nearly scale-invariant for all \hbox{$m > 2$}, with
\hbox{$0.93 < n_s < 0.97$}. For the case \hbox{$m = 4$}, the density
fluctuation amplitude is independent of \hbox{$\left(\mu / m_{Pl}\right)$},
and there is no intrinsic lower bound on $\mu$ from inflationary constraints.
For \hbox{$m > 4$}, the density fluctuation amplitude $\delta$ {\it
decreases} with decreasing \hbox{$\left(\mu / m_{Pl}\right)$}, and no
fine-tuning of $\Lambda$ is required. Potentials which contain quadratic
terms can still be dominated by terms of order \hbox{$m > 2$} if the
condition (\ref{socondition}) on the second order slow-roll parameter is met.
This allows placement of a lower limit on \hbox{$\left(v / m_{Pl}\right)$}.

These results are illustrated by a model in which the inflationary potential
is generated by gauge boson loop effects in a Lagrangian with an explicitly
broken ${\rm SO(3)}$ symmetry. In this model, the potential is dominated by
terms of order \hbox{$m = 4$}, but its fourth derivative, describing the
scalar particle self-coupling, diverges logarithmically at the origin.
However, the general analysis is valid up to logarithmic corrections to the
limit on the coupling constant obtained from the COBE observation
\hbox{$\delta \simeq 10^{-5}$}:
\begin{eqnarray}
g^4 \ln\left(g^2\right) =&& g_0^4 \ln\left(g_0^2\right) \left(1 +
{\ln\left[\left(\pi / 60\right) \left(v / m_{Pl}\right)^2\right]  \over
\ln\left(g_0^2\right)}\right)^{-1}\cr
g_0^4 \ln\left(g_0^2\right) =&& -{8 \over 3} \left(\pi^2 \over 15\right)^3
\delta^2,
\end{eqnarray}
where \hbox{$g \simeq 10^{-3}$} for a wide range of scales $v$. The spectral
index $n_s$ is given exactly by the value derived in the general analysis,
\hbox{$n_s = 0.95$}, independent of the symmetry breaking scale. Couplings to
fermionic sectors are discussed. In a model with fermions, in which the
Yukawa couplings also break the ${\rm SO(3)}$ symmetry, quadratic terms are
introduced into the potential, and a lower bound on \hbox{$\left(v /
m_{Pl}\right)$} is obtained:
\begin{equation}
\left(v \over m_{Pl}\right) > \left(h \over g\right)^2 \sqrt{160
\ln\left(h^2\right) \over \pi \ln\left(g^2\right)},
\end{equation}
where h is the Yukawa coupling to a neutral fermion. For weakly coupled
fermions, \hbox{$h \simeq 10^{-6}$}, inflation is consistent with a symmetry
breaking scale \hbox{$v \simeq m_{GUT} \simeq 10^{16} {\rm GeV}$}. Models
with ${\rm SO(3)}$ symmetric fermion sectors possess no such lower limit.

These results in many respects do not bode well for efforts at
``reconstruction'' -- the determination of specific details -- of the
inflationary potential from accurate measurement of the scalar spectral index
and tensor fluctuation amplitude
\cite{copeland93,ho90,sa92,lu92,li92,da92,co93l,tu93,co94,li94,ad95}. For a
large class of viable models, the tensor fluctuation amplitude is very small,
and other cosmological constraints are largely insensitive to the specific
form of the scalar field potential, leaving little opportunity to distinguish
one model from another using cosmological observations alone.

\section*{Acknowledgements}

The authors would like to thank S.P. deAlwis for helpful discussions. This
work is supported in part by U.S. Department of Energy Grant No.
DEFG-ER91-406672.

\end{document}